\documentclass[preprints,article,accept,pdftex,moreauthors]{Definitions/mdpi} 
\firstpage{1} 
\pubvolume{1}
\issuenum{1}
\articlenumber{0}
\pubyear{2022}
\copyrightyear{2022}
\datereceived{} 
\dateaccepted{} 
\datepublished{} 
\hreflink{} 
\pdfoutput=1

\usepackage{amsmath}
\usepackage{amssymb}
\usepackage{amstext}
\usepackage{bm}
\usepackage{xcolor}
\usepackage{helvet}
\usepackage{graphicx} 
\usepackage{physics}
\usepackage{qcircuit}

\Title{Correcting Coherent Errors by Random Operation on Actual Quantum Hardware}

\TitleCitation{Correcting Coherent Errors by Random Operation on Actual Quantum Hardware}

\Author{Gabriele Cenedese $^{1,2}$, Giuliano Benenti $^{1,2,3}$\orcidB{} and Maria Bondani $^{4,*}$\orcidC{}}

\AuthorNames{Gabriele Cenedese, Giuliano Benenti and Maria Bondani}

\AuthorCitation{Cenedese, G.; Benenti, G.; Bondani, M.}

\address{%
$^{1}$ \quad Center for Nonlinear and Complex Systems, Dipartimento di Scienza e Alta Tecnologia, Universit\`a degli Studi dell'Insubria, via Valleggio 11, 22100 Como, Italy \\
$^{2}$ \quad Istituto Nazionale di Fisica Nucleare, Sezione di Milano, via Celoria 16, 20133 Milano, Italy\\
$^{3}$ \quad NEST, Istituto Nanoscienze-CNR, I-56126 Pisa, Italy\\
$^{4}$ \quad Istituto di Fotonica e Nanotecnologie, Consiglio Nazionale delle Ricerche, via Valleggio 11, 22100 Como, Italy
}

\corres{Correspondence: maria.bondani@uninsubria.it}

\abstract{Characterizing and mitigating errors in current noisy intermediate-scale devices is important to improve performance of next generations of quantum hardware. In order to investigate the importance of the different noise mechanisms affecting quantum computation, we perform full quantum process tomography of single qubits in a real quantum processor in which echo experiments are implemented. Besides error sources already included in standard models, the obtained results show the dominant role of coherent errors, which we practically correct by inserting random single-qubit unitaries in the quantum circuit, significantly increasing the circuit length over which  
quantum computations on actual quantum hardware produce reliable results.}


\begin{document}


\section{Introduction}

\label{sec:intro}

Quantum computers operating with $\sim$ 50-100 qubits may be able to perform tasks which surpass the capabilities of today's classical digital super-computers~\cite{qcbook,Preskill2018}, 
and the quantum advantage for specific problems has been recently claimed~\cite{Martinis2019,Pan2020,Zoller2022}.
However, quantum advantage can only be reached with a high enough quantum gate precision and through processes generating a large enough amount of entanglement, able to outperform classical tensor network methods~\cite{Waintal2020}. 
Unfortunately, present-day Noisy Intermediate-Scale Quantum (NISQ) devices suffer from significant decoherence and the effects of various noise sources such as residual inter-qubit coupling or interactions with an uncontrolled environment. Noise limits the size of quantum circuits that can be executed reliably and, therefore, achieving the quantum advantage in complex, practically relevant problems is still an imposing challenge. 

It is then important to benchmark the progress of currently available quantum computers~\cite{Gambetta19,Benenti21}, and possibly find suitable error mitigation strategies.
Here, we focus on freely available IBM quantum processors. For them, 
IBM provides a few noise parameters characterizing: qubit relaxation time, 
qubit dephasing time, error rates in single-qubit gates, in two-qubit CNOT gates,
and in quantum measurements. Such parameters are updated after each hardware calibration and provide very useful information to assess the performance of quantum hardware.
Nevertheless, as we will show below, these noise channels are not 
sufficient for an accurate description of the errors affecting a quantum computer.
As a general observation, we note that even in the simplest case of memoryless errors, the general description of quantum noise in terms of quantum operations requires a large number of real parameters $N_p = N^4-N^2$, where $N=2^n$ is the Hilbert space dimension for $n$ qubits~\cite{qcbook}.
For a single qubit $N_p = 12$, and these parameters have a simple intuitive interpretation in terms of rotations, deformations, and displacements of the Bloch sphere. 
Furthermore, coherent errors~\cite{Dima00,Dima00bis,Benenti01,Benenti01bis,Benenti03,Montangero03,Cory06,Benenti07,Greenbaum18,Majumder22}, that is, unitary errors that are slowly varying relative to the gate time, can arise due to several reasons, including  miscalibration or drift out of calibration of the control system used to drive the qubit operations, cross-talk with neighboring qubits, external fields, and residual qubit–qubit interactions. Such errors cannot be removed with standard 
error correcting codes~\cite{qcbook} developed for stochastic (incoherent), 
uncorrelated, memoryless errors.

Here we describe the evolution of a qubit inside an operating quantum computer as a quantum operation, or a completely positive trace-preserving (CPT) map, acting on the single qubit Bloch sphere. Our first goal is to provide a full characterization of such CPT map, that is, a \textit{quantum process tomography}, going beyond the few above mentioned noise channels, in order to analyze the performances of current noisy quantum hardware and define a useful benchmark for new releases of quantum computers. As the quantum noise channel we consider 
a \textit{quantum echo experiment}, reversing a quantum computation, so that 
in the ideal, noiseless case we would reconstruct the initial state.
More specifically, we consider an even sequence of CNOT gates 
($\hbox{CNOT}^2=I$) and a more general sequence of two-qubit random unitary operators, combined with their time-reversal operators to perform the echo experiment.  
The comparison between the 
results from the quantum noise tomography on the IBM quantum hardware 
and those obtained with the Qiskit simulator, which only takes into account a limited number of noise channels, highlight the relevance of coherent errors.
This observation suggests the introduction of random single-qubit gates 
into the logical circuit, to suppress coherent errors~\cite{Alber05,Emerson16}.
We demonstrate the effectiveness of such a strategy in the above
CNOT-echo experiment.

The paper is organized as follows. In Section~\ref{sec:Bloch} we recall the Bloch sphere representation of a CPT map for a single qubit,  
and the main steps for the consequent quantum process tomography. 
In Section~\ref{sec:results}
echo results obtained from IBM quantum processors are shown and compared with those from the IBM simulator. In Section~\ref{sec:correction}
we discuss the effectiveness of 
the randomization strategy for coherent error correction. 
Our conclusions are finally drawn in Section~\ref{sec:conc}.

\section{CPT maps on the Bloch sphere}

\label{sec:Bloch}

As it is well known, a two-level system (a qubit) state can be represented by a point in a ball of unit radius, called the Bloch ball, embedded in $\mathbb{R}^3$, which defines the so-called Bloch vector ($\bm{r}$). Pure states are those that lie on the surface of the ball ($|\bm{r}|=1$), i.e. the Bloch sphere, while mixed states are those inside the sphere ($|\bm{r}|<1$). An ideal quantum echo experiment should leave pure states on the sphere, while in the real case the length of the Bloch vector in general decreases.

Given an initial single-qubit state $\rho$, we consider the quantum noise channel, or completely positive trace-preserving (CPT) map $\mathbb{S}$, as quantum black box, representing the qubit interacting with a generic physical system. Without any a-priori knowledge of the quantum noise processes affecting the qubit, we can reconstruct $\mathbb{S}$ by preparing different input states $\rho$ and for each of them measuring the output state $\rho' = \mathbb{S}(\rho)$. In the Bloch-sphere representation, the Bloch vector evolves as an affine map:
\begin{linenomath}
	\begin{equation}
	\bm{r} \rightarrow \bm{r}'=M\bm{r}+ \bm{c}.
	\end{equation}
\end{linenomath}
To perform a full quantum process tomography, i.e. to reconstruct the matrix $M$ and the vector $\bm{c}$, we need to perform 12 different experiments. In each experiment, we
prepare one out of the four initial states 
\begin{equation}
\ket{0}, \;\;\ket{1},\; \;\ket{x}=\frac{1}{\sqrt{2}}(\ket{0}+\ket{1}), \;\;\ket{y}=\frac{1}{\sqrt{2}}(\ket{0}+i\ket{1}).
\label{eq:psiin}
\end{equation}For each initial state, down the channel we measure the
polarization of the final state $\rho'$ along one of the three coordinate axes, in order to estimate the final Bloch vector for each input state. Each experiment is repeated a large number of times, $N_r=25$, with $N_m=8192$ runs each time, to reconstruct $M$ and $\bm{c}$ with high accuracy. In the case of a sequence of random unitaries, $N_r=25$ sequences of two-qubit random
unitaries are extracted from the Haar measure on the unitary group $U(4)$~\cite{Pozniak98,Weinstein05,Vahid22}.

The quality of the quantum channel is measured by the fidelity $\mathcal{F}$ between the ideally pure 
single-qubit initial state $\ket{\psi_{in}}$ and the final, generally mixed state $\rho_{out}$:
\begin{linenomath}
	\begin{equation}
\mathcal{F}=\bra{\psi_{in}}\rho_{out}\ket{\psi_{in}}.
	\end{equation}
\end{linenomath}
We will evaluate the fidelity, both with Qiskit simulator and on the real quantum hardware, 
for the initial states of Eq.~(\ref{eq:psiin}) and the echo protocols described below.


\section{Single-qubit quantum process tomography}
All experiments on actual quantum hardware were performed on $ibm\_lagos$, in particular, we reconstructed the evolution of the Bloch sphere of qubit 0 ($q_0$) with qubit 1 ($q_1$) as the ancilla. The CNOT gates of the CNOT noisy channel were performed with $q_0$ as the control qubit. Note that using $q_0$ as the target qubit does not qualitatively alter the results.
\label{sec:results}

\subsection{CNOT noisy channel}
\begin{figure}[H]
	\includegraphics[width=15 cm]{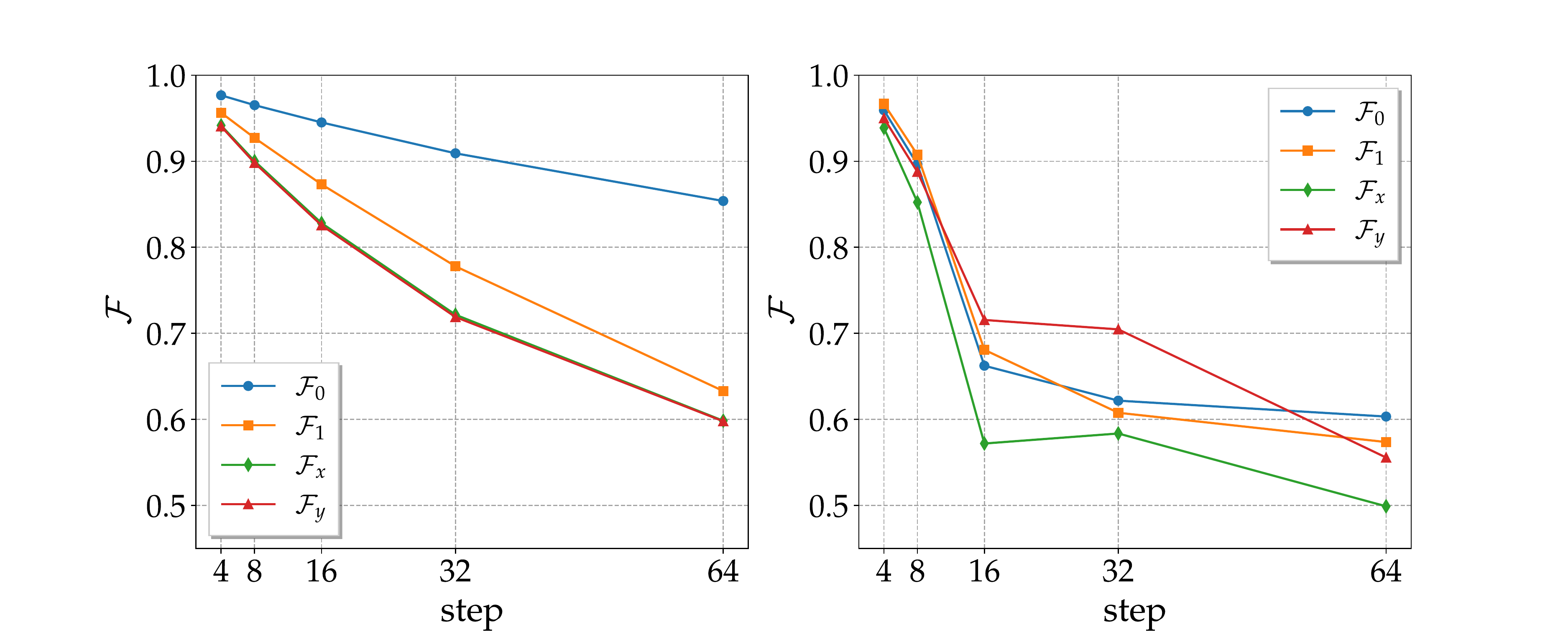}
	\caption{Fidelities of the 4 states used to reconstruct the CPT map as a function of the 
	number of steps (one step of the CNOT noisy channel corresponds to two CNOT gates). Left: simulations with the noise parameters of \textit{ibm\_lagos}, calibration of November 2nd, 2022. Right: actual results obtained with \textit{ibm\_lagos} on November 2nd, 2022.
	\label{fig:fidelityCNOT}}
\end{figure}

\begin{figure}[H]
	\begin{adjustwidth}{-\extralength}{0cm}
		\centering
		\includegraphics[width=18cm]{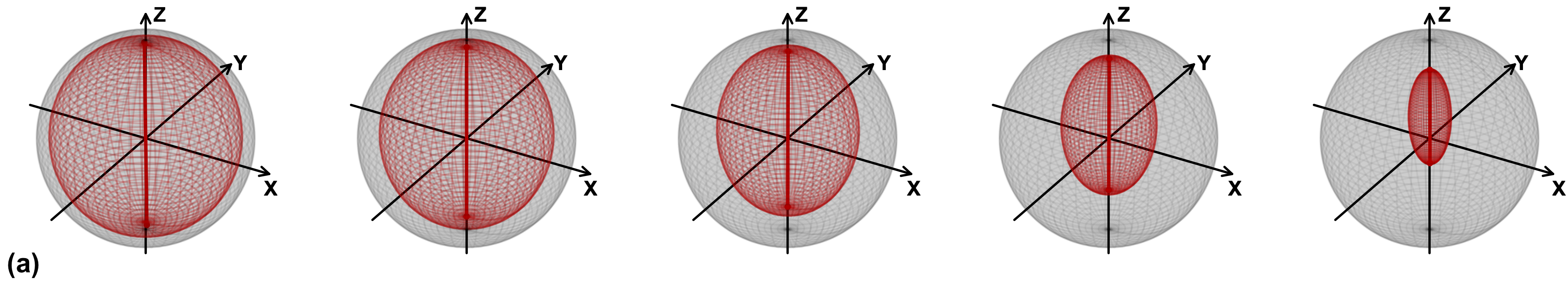}
		\includegraphics[width=18cm]{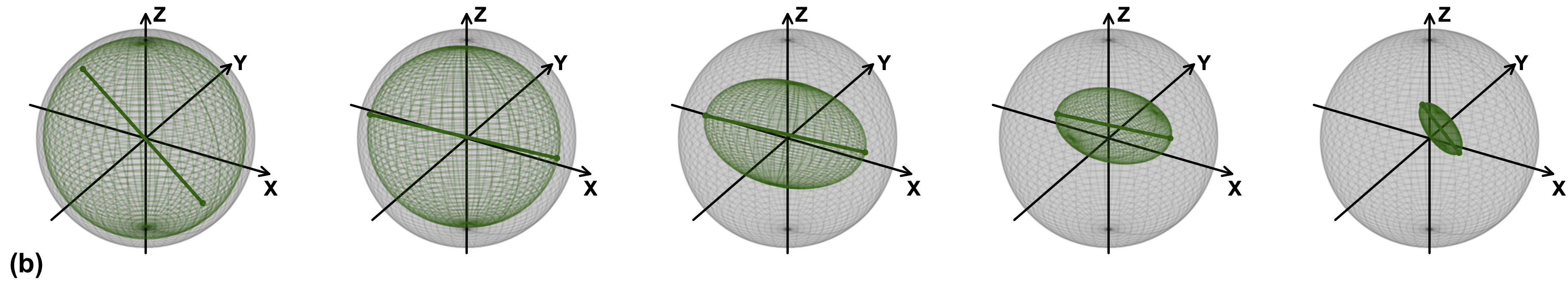}
	\end{adjustwidth}
	\caption{Evolution of the single qubit Bloch ball as a function of 
	the number of CNOT map  steps. 
	From left to right the number of steps increases, in correspondence with the data shown in 
	Figure~\ref{fig:fidelityCNOT}. (\textbf{a}): Qiskit Simulations with the noise parameters of \textit{ibm\_lagos}. The segment highlighted in red shows the major axis of the ellipsoid, the gray sphere is the unit-radius Bloch ball. (\textbf{b}): Results obtained with \textit{ibm\_lagos}.
	The segment highlighted in green shows the major axis of the ellipsoid, the gray sphere is unit-radius Bloch ball.
	Data from the quantum processor taken on 2nd November 2022, with the corresponding calibration parameters used for Qiskit simulations.} 
	\label{fig:sphereCNOT}
\end{figure}

The fidelities obtained with the sequence of noisy CNOTs (in short, CNOT noisy channel) are shown in Figure~\ref{fig:fidelityCNOT}. Results from the Qiskit simulations are easily interpreted as follows. Since one problem of the CNOT is the long gate time, the dephasing and the energy relaxation errors (parametrized by the decoherence times $T_1$ and $T_2$) become important. The state $\ket{0}$, which is not affected by these kind of errors, present better fidelities than $\ket{1}$, $\ket{x}$ and $\ket{y}$; with $\ket{1}$ slightly better than $\ket{x}$ and $\ket{y}$, since it is only affected by the relaxation. The results obtained with the actual device do not reflect the Qiskit simulation, and the reason can be understood by looking at the evolution of the Bloch sphere depicted in Figure~\ref{fig:sphereCNOT}. In the Qiskit simulations the sphere deforms to become an ellipsoid with semi-major axis at the $z$-axis,  and center shifted toward the direction of positive $z$'s. This mirrors what happens with fidelities, with states near the north pole (which represent the state $\ket{0}$) of the spheres being gradually less affected by dephasing and energy relaxation. In the real 
quantum hardware case, ellipsoids are still formed, but they appear to be rotated, with the semi-major axis no longer in the $z$-direction. This fact can be interpreted 
(see below) as the result of some kind of coherent errors, which 
induce undesired rotations of the Bloch sphere.

\subsection{Random unitaries}
\begin{figure}[H]
	\includegraphics[width=15 cm]{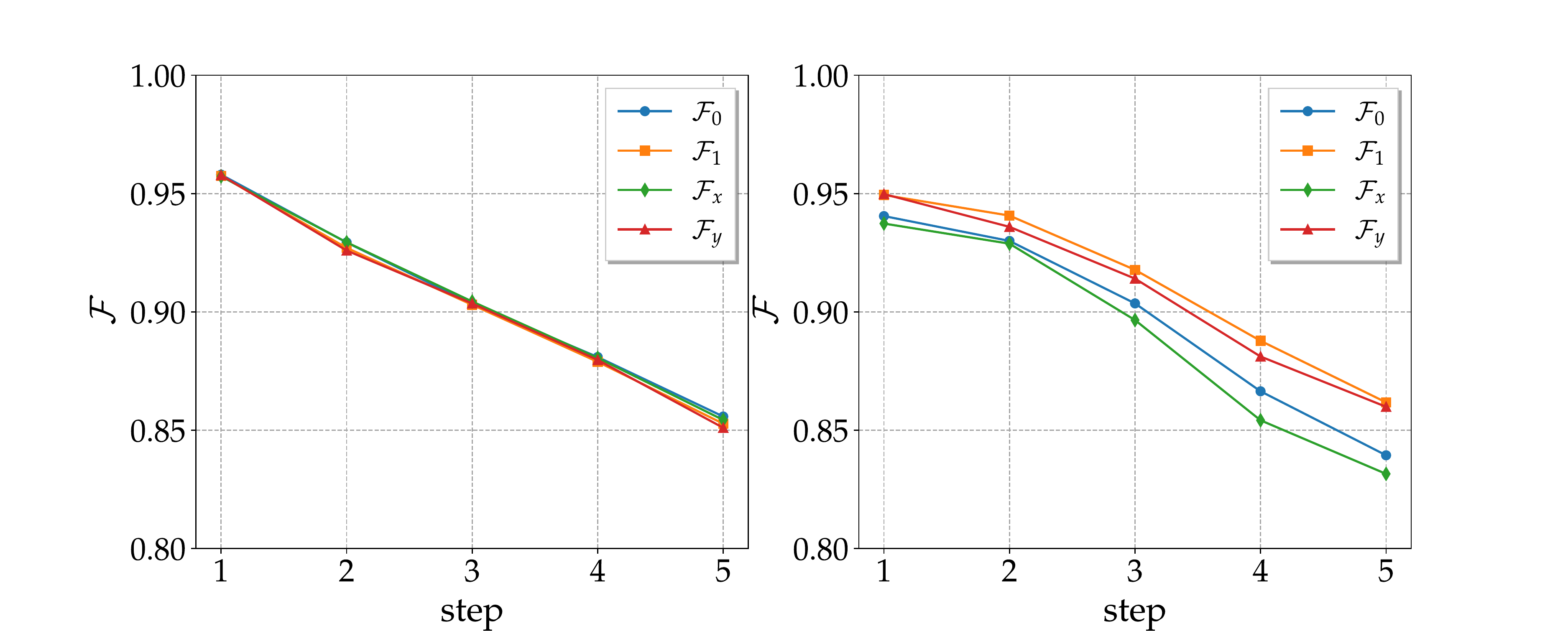}
	\caption{Same as in Figure~\ref{fig:fidelityCNOT}, but for random unitaries.
    Qiskit (left) and actual hardware (right) data were 
    obtained with \textit{ibm\_lagos} on 2nd November 2022.}
	 \label{fig:fidelityRANDOM}
\end{figure}   

\begin{figure}[H]
	\begin{adjustwidth}{-\extralength}{0cm}
		\centering
		\includegraphics[width=18cm]{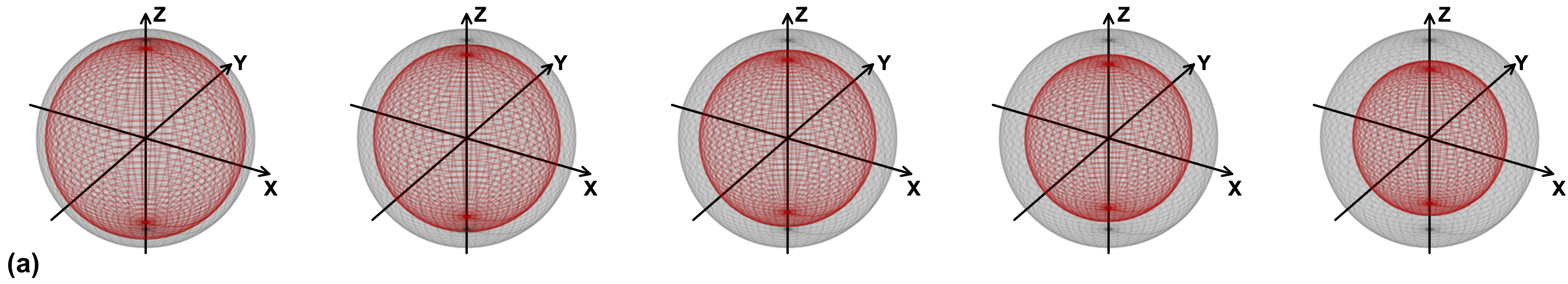}
		\includegraphics[width=18cm]{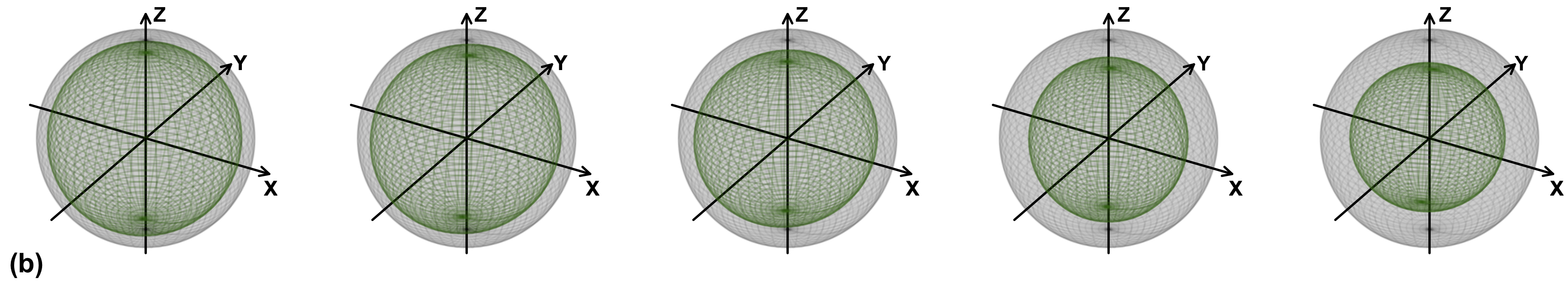}
	\end{adjustwidth}
	\caption{
	Same as in Fig.~\ref{fig:sphereCNOT}, but for the random unitaries channel of Fig.~\ref{fig:fidelityRANDOM}}
	\label{fig:sphereRANDOM}
\end{figure}

We now consider sequences of two-qubit random unitaries $U_k$, 
with $k$ running from 1 to the number of steps $N_s$, combined with 
their inverse $U_k^\dagger$ to realize the echo experiment. 
To summarize, we implement the overall random quantum circuit
$\prod_{k=1}^{N_s} U_k^\dagger U_k$.
In contrast with the CNOT channel, as we can see in Figure~\ref{fig:fidelityRANDOM}, Qiskit simulation's fidelities of the four basis states are comparable. This can be easily understood: the CNOTs in the decomposition into elementary quantum logic gates of general two-qubit unitary operators are always preceded by a single qubit random rotation (see appendix~\ref{sec:KAK}).
As a consequence, whatever the initial state is, CNOTs act on a random state of the sphere, and the 
error due to dephasing and relaxation is on average independent of the initial state. 
For the same reason, we anticipate that the effects of coherent errors cancel.
Indeed, we expect from previous literature~\cite{Alber05} that 
the deleterious effect of such errors is greatly reduced if they are randomized
by repeatedly rotating the computational basis via random single-qubit unitaries.
The fact that, unlike the CNOT channel, the actual hardware has comparable performance
to that predicted by the Qiskit simulator (see Figure~\ref{fig:sphereRANDOM}), 
demonstrates that for the CNOT channel hardware peformance degradation
should be ascribed to a large extent to coherent errors.

 \section{Error correction by randomization}

\label{sec:correction}

\begin{figure}[H]
	\includegraphics[width=15 cm]{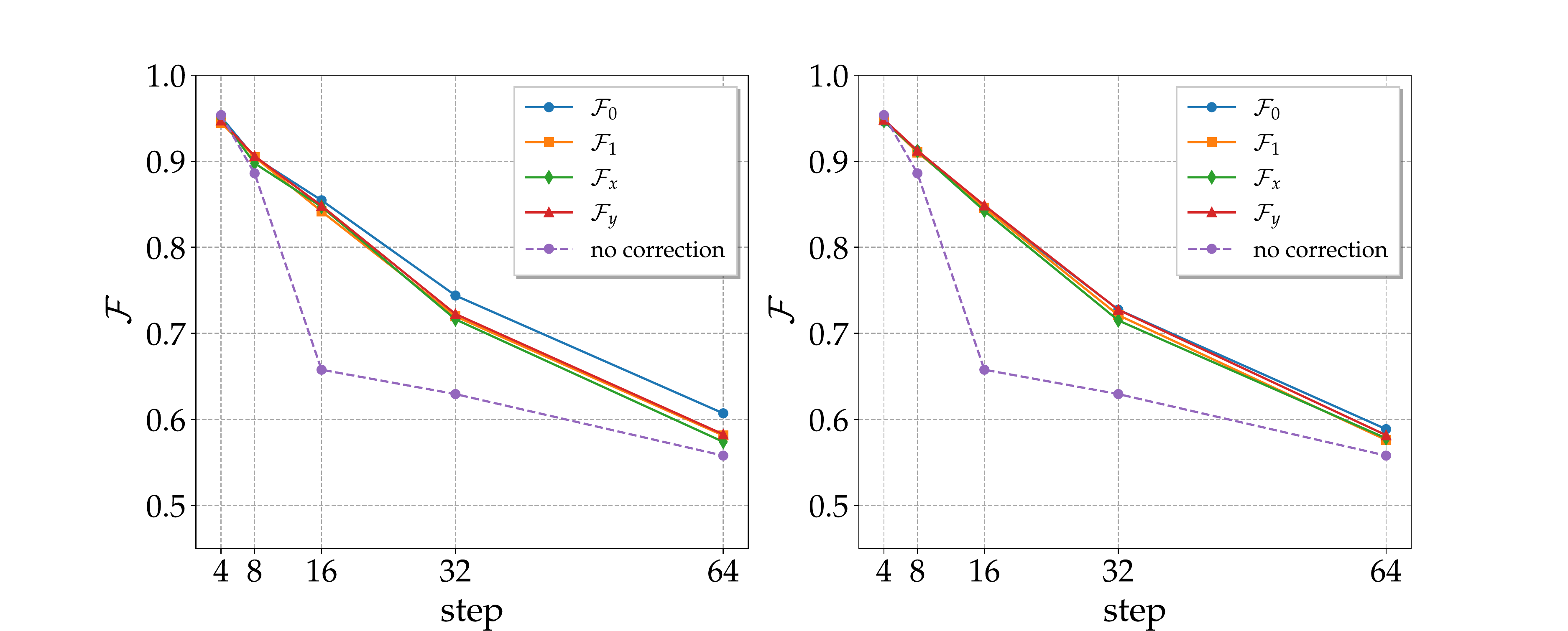}
	\caption{Fidelities of the 4 basis states as a function of the number of
steps (one step of the CNOT noisy channel corresponds to two CNOT gates) after the correction procedure described above. On the left computational basis is rotated with random unitaries, on the right with single-axis rotations. The purple curve represents the average fidelities obtained without correction. Results obtained with $ibm\_lagos$ on 20th November 2022.\label{fig:fidelityCorrection}}
\end{figure}

 In the case of the CNOT noisy channel the nature of the coherent noise suggests a natural error correction procedure via repeatedly rotating the computational basis. The idea is, before any CNOT pair, to apply a random single qubit unitary gate at each qubit of the noise channel and the respective adjoint after the CNOTs, in order to globally obtain the identity. Alternatively, we took randomly chosen rotations with respect to x or y (picked randomly as well), in order to decrease the number of logic gates used for the correction. As we can see in Figure~\ref{fig:fidelityCorrection} the fidelities significantly improve when compared with those of the CNOT noisy channel without correction and, moreover, they are independent of the initial state. The results with single-axis rotations are comparable with those of generic single-qubit unitaries operators.




\section{Conclusions}

\label{sec:conc}

We have pointed out the presence of coherent errors through echo experiments performed on actual quantum hardware. In particular we have considered two quantum noise channels, one formed by a sequence of CNOT gates, the other by random two-qubit interactions. The former clearly shows the presence of coherent errors by observing the rotation of Bloch spheres. The latter suggests that a randomization of the computation basis leads to a natural cancellation of coherent errors,
as we practically demonstrate for a sequence of CNOT gates. 
Getting rid of this kind of noise is the first step in order to implement a successful error correction procedure.

\vspace{6pt} 


\authorcontributions{G. C. performed quantum simulations by coding actual IBM quantum processors. G. B. and M. B. supervised the work. All authors discussed the results and contributed to writing and revising the manuscript.}

\funding{G. C. and G. B. acknowlewdges the financial support of the INFN through the project QUANTUM.}

\institutionalreview{Not applicable.}

\informedconsent{Not applicable.}

\dataavailability{The dataset used and analyzed in the current study are available from the corresponding author on reasonable request.} 

\acknowledgments{We acknowledge use of the IBM Quantum Experience for this work. The views expressed are those of the authors and do not reflect the official policy or position of IBM company or the IBM-Q team.}

\conflictsofinterest{The authors declare no conflict of interest. The funders had no role in the design of the study; in the collection, analyses, or interpretation of data; in the writing of the manuscript, or in the decision to publish the results.}





\appendixtitles{yes} 
\appendixstart
\appendix
\section[\appendixname~\thesection]{Cartan's KAK decomposition of the unitary group}
\label{sec:KAK}
\begin{figure}[H]
    \centering
	\includegraphics[width=11 cm]{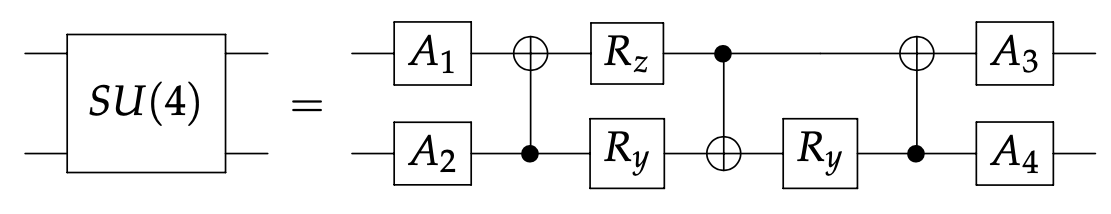}
	\caption{A quantum circuit implementing a two qubits unitary gate using the KAK parametrization of $SU(4)$.}  
	\label{fig:KAK}
\end{figure}

The Cartan's KAK decomposition can be used for constructing an optimal quantum circuit for achieving a general two-qubit quantum gate, up to a global phase, which requires at most 3 CNOT and 15 elementary one-qubit gates from the family $\{R_y,R_z\}$, i.e. single-qubit rotations obtained by exponentiating the corresponding Pauli matrices. It can be proved that this construction is optimal, in the sense that there is no smaller circuit, using the same family of gates, that achieves this operation \cite{vatan2004optimal}. 
Following the general prescription \cite{khaneja2001cartan,khaneja2001time}
one can decompose every $SU(4)$ element as depicted in Figure~\ref{fig:KAK}, where $A_j\in SU(2)$ are single qubit unitaries decomposable into elementary one-qubit gates according to the well-known Euler strategy.

Note that in order to randomly extract one of these operators the angles of the single qubit rotations must be extracted uniformly with respect to the Haar measure of the unitary group.

\begin{adjustwidth}{-\extralength}{0cm}

\reftitle{References}



\externalbibliography{yes}
\bibliography{main.bib}

\end{adjustwidth}
\end{document}